

\input phyzzx
\overfullrule=0pt
\baselineskip=18pt
\hfill {IP-ASTP-12}\break
\FRONTPAGE
\strut\hfill { hep-th/9506093} \vglue 0.8in
\font\twelvebf=cmbx12 scaled\magstep2

\vskip 0.0in
\centerline {\twelvebf Generalizing the Coleman Hill Theorem}

\vskip .5in
\centerline{{\it Hsien-Chung Kao}}
\vskip .1in
\centerline{ Institute of Physics, Academia Sinica}
\centerline{ Nankang, Taipei, 11529 Taiwan}
\vskip 0.4in
\centerline{\bf ABSTRACT}
\vskip 0.1in

Following the work of Khare {\it et al}, we show that the generalization to
systems with spontaneous symmetry breaking of the Coleman-Hill theorem to
one-loop order, can be extended to the case including fermions with the most
general interactions.  Although the correction to the parity-odd part of the
vacuum polarization looks complicated in the Higgs phase, it turns out that
the correction to the Chern-Simons term is identical to that in the symmetric
phase, with the difference coming only from the contribution of the would be
Chern-Simons term.  We also discuss the implication of our result to
nonabelian systems.

\vfill

\footnote{}{Email Address: hck$@$phys.sinica.edu.tw}

\vfill \endpage

\def\pr#1#2#3{Phys.  Rev.  {\bf D#1}, #2 (19#3)}
\def\prl#1#2#3{Phys. Rev.  Lett.  {\bf #1}, #2 (19#3)}

\def\np#1#2#3{Nucl.  Phys.  {\bf B#1}, #2 (19#3)}
\def\pl#1#2#3{Phys.  Lett.  {\bf B#1}, #2 (19#3)}
\def\ibid#1#2#3{ {\it ibid.} {\bf #1}, #2 (19#3)}

\def\tr{\,{\rm Tr}}

\def\hp{\hat{\phi}}
\def\ha{\hat{A}}
\def\tvp{\tilde{\varphi}}
\def\vp{\varphi}


\REF\rCSFT{J. Schonfeld, \np{185}{157}{81}; S. Deser, R. Jackiw and S.
Templeton, Ann. Phys. (N.Y.) {\bf 140}, 37 (1982).}

\REF\rCSANYON{D. p. Arovas, J. R. Schrieffer, F. Wilczek and A. Zee,
\np{251}{117}{85}; T. H. Hnsson, M. R\u ocek, I. Zahed and S. C. Zhang,
\pl{214}{475}{88}.}

\REF\rFQHE{R. E. Prange and S. M. Girvin, {\it The Quantum Hall
Effect}, (Springer-Verlag, Berlin, 1990).}

\REF\rHong{ J.  Hong, Y.  Kim and P.Y.  Pac, \prl{64}{2230}{90}; R.
Jackiw and E.J.  Weinberg, \ibid{64}{2234}{90}; R.  Jackiw, K.  Lee
and E.J.  Weinberg, \pr{42}{3488}{90}; Y.  Kim and K.  Lee,
\ibid{49}{2041}{94}.}

\REF\rColeman{S.  Coleman and B.  Hill, \pl{159}{184}{85}.}

\REF\rSemenoff{G.W.  Semenoff, P.  Sodano and Y.-S.  Wu,
\prl{62}{715}{89}; W.  Chen, \pl{251}{415}{91}; V.P.  Spiridonov and
F.V.  Tkachov, \ibid{260}{109}{91}; D.K. Hong, T. Lee and S.H. Park,
\pr{48}{3918}{93}.  } 

\REF\rKhleb{S.Y.  Khlebnikov, JETP letters {\bf 51}, 81, (1990); V.P.
Spiridonov.  \pl{247}{337}{90}.  } 

\REF\rCHTh{A. Khare, R. MacKenzie, and M. B. Paranjape, \pl{343}{239}{95}.}

\REF\rPisarski{R.D.  Pisarski and S.  Rao, \pr{32}{2081}{85}; G.
Giavirini, C.P.  Martin and F.  Ruiz Ruiz, \np{381}{222}{92}.}

\REF\rKhlebn{S.Y.  Khlebnikov and M.E.  Shaposhnikov, \pl{254}{148}{91}.}

\REF\rDunne{L. Chen, G. Dunne, K. Haller, E. Lim-Lombridas, \pl{348}{468}{95}.}

\REF\rKhare{A. Khare, R. B. MacKenzie, P. K. Pannigrahi, and M. B. Paranjape,
``Spontaneous Symmetry Breaking and the Renormalization of the Chern-Simons
Term'', UdeM-LPS-TH-93-150 (1993), hep-th/9306027, accepted for publication in
Phys. Lett. B.}

\REF\rCscf{H-C. Kao, K. Lee, C. Lee and T. Lee, \pl{341}{181}{94}.}

\REF\rJackiw{R. Jackiw, \pr{9}{1686}{73}.}

\REF\rDeri{I.J.R. Aitchison and C.M. Fraser, \pl{146}{63}{84}.}

\REF\rCSII{C.  Lee, K.  Lee, and E.J.  Weinberg, \pl{243}{105}{90};
E.A.  Ivanov,\ibid{268}{203}{91};
S.J.  Gates and H.  Nishino, \ibid{281}{72}{92}.}

\REF\rCSIII{H.-C.  Kao and K.  Lee, \pr{46}{4691}{92}.}

It is known that the Chern-Simons theories can give rise to particle
excitations with fractional spin and statistics, and are thus relevant to the
fractional quantum Hall effect [\rCSFT, \rCSANYON, \rFQHE].  Further studies
show that this property is also enjoyed by the topological vortices in the
Higgs phase of these systems [\rHong].  Since the inverse of the Chern-Simons
coefficient plays the role of statistical parameter, the knowledge of its
quantum correction is important for a complete understanding of the quantum
physics in these systems.

In the absence of massless charged particles and spontaneous symmetry
breaking, Coleman and Hill have shown that the only correction to the
Chern-Simons coefficient comes from the fermion one-loop contribution
[\rColeman].  When the two conditions are not satisfied, higher-loop effect
is generally non-vanishing and the corrections are complicated functions of
couplings and particle masses [\rSemenoff, \rKhleb].  By analyzing the
one-loop correction for a system without fermion, Khare {\it et al} show
that in terms of the effective action the Chern-Simons term does not get
renormalized even in the Higgs phase [\rCHTh].  This suggests that the above
theorem can be generalized to systems with spontaneous symmetry breaking if
recast in terms of the effective action.

When the gauge symmetry is nonabelian, the coefficient must be quantized
for the system to be quantum-mechanically consistent.  In the symmetric phase,
this has been explicitly verified to one-loop [\rPisarski].  In the Higgs
phase, the situation is more subtle.  If the gauge symmetry is completely
broken, since there is no well-defined symmetry generator, we do not expect
the Chern-Simons coefficient there to be quantized [\rKhlebn].
On the other hand, if there is remaining symmetry in the Higgs phase,
we do believe and it has be shown that the corresponding Chern-Simons
coefficient satisfy the quantization condition [\rDunne, \rKhare].

In this letter, we extend the result in [\rCHTh] to systems containing
also a fermion.  Using the background field method, we calculate the
coefficient of the would be Chern-Simons term to one loop.  In terms of the
effective action, we show that the one-loop correction to the Chern-Simons
term in the Higgs phase is the same as that in the symmetric phase, with the
difference coming solely from the contribution of the would be Chern-Simons
term.  We speculate that similar situation happens in the nonabelian case
so that all the corrections to the Chern-Simons coefficients are identical,
if we subtract out the contribution from the would be Chern-Simons term.

Let us consider a model with a gauge field $A_\mu$, a complex
Higgs field $\phi$, and a Dirac fermion field $\psi $.  The most
general gauge-invariant renormalizable Lagrangian is given as
$$
\eqalign{ {\cal L} =
& - {1\over 4e^2} F_{\mu\nu}^2 + {\kappa \over 2}
\epsilon^{\mu\nu\rho}A_\mu \partial_\nu A_\rho + |D_\mu \phi|^2
+ i\bar{\psi}\gamma^\mu D_\mu \psi \cr
& -(m^2 |\phi|^2 + {4\lambda \over 4!}|\phi|^4 + {8\tau \over 6!}|\phi|^6)
- (M + 2 g_1 |\phi|^2 )\bar{\psi}\psi
- g_2 [\phi^2 \bar{\psi}\psi^* + \phi^{*2} \bar{\psi}^*\psi]. \cr}
\eqno\eq
$$
where $D_\mu = \partial_\mu + iA_\mu$ and all coupling constants are
real [\rCscf].  The metric and the gamma matrices are chosen to be
$\eta_{\mu\nu} = {\rm diag}(1,-1,-1,),$ and
$\gamma^\mu= (\sigma^2, i\sigma^3, i\sigma^1 )$
so that the gamma matrices satisfy
$\gamma^\mu \gamma^\nu = \eta^{\mu\nu} -i\epsilon^{\mu\nu\rho}
\gamma_\rho$, with $\epsilon^{012} = \epsilon_{012} = 1$.

For later convenience, we express the scalar and spinor fields
in terms of real and Majorana fields:
$\phi = (\phi_1 + i \phi_2)/\sqrt{2}$ and
$\psi = (\psi_1 + i\psi_2)/\sqrt{2}$.
To proceed, we separate $(\phi_a, A_\mu)$ into the background part
$(\hp_a, \ha_\mu)$ and the quantum part $(\phi_a, A_\mu)$.  The background
Lagrangian is then given by
$$
{\cal L}_B = {\cal L}(\ha + A, \hp + \phi, \psi)
- {\cal L}(\ha, \hp, 0)
- \phi_a {\partial {\cal L} (\ha,\hp,0) \over \partial \hp_a}
- A_\mu {\partial {\cal L} (\ha,\hp,0) \over \partial \ha_\mu}
\eqno\eq
$$
In the background $R_\xi$ gauge,
$$
{\cal L}_{gf} = -{1\over 2\xi}
(\partial^\mu A_\mu + \xi \hp \times \phi)^2, \eqno\eq
$$
which gives rise to the Fadeev-Popov ghost Lagrangian
$$
{\cal L}_{FP}
= \bar{\eta} \left\{
-\partial_\mu^2 - \xi (\hp^2 - \hp \cdot \phi)\right\}\eta. \eqno\eq
$$
Here,
$$
\eqalign{
& \hp \times \phi \equiv \epsilon_{ab} \hp_a \phi_b \cr
& \hp \cdot \phi \equiv \hp_a \phi_a. \cr} \eqno\eq
$$

Combining the Lagrangian (2) and the gauge fixing terms (3) and (4),
we see to quadratic term
$$
\eqalign{ {\cal L}_B =
& {1\over 2} A^\mu U^{\mu\nu}(\hp) A^\nu + A^\mu V_{\mu a}(\hp,\ha) \phi_a
+ {1\over 2} \phi_a W_{ab}(\hp,\ha) \phi_b \cr
& + {1\over 2} \bar{\psi}_a T_{ab}(\hp,\ha) \psi_b
+ \bar{\eta} R(\hp) \eta, \cr}\eqn\LBq
$$
with
$$
\eqalign{
& U_{\mu\nu}(\hp) =
\left[-P^2/e^2 + \hp^2\right]\eta_{\mu\nu}
+ \left[-{1\over \xi} + {1\over e^2}\right] P_\mu P_\nu
+ i\kappa \epsilon_{\mu\nu\rho} P^\rho \cr
& V_{\mu a}(\hp,\ha) =
-2\epsilon_{ab}(\partial_\mu\hp_b) - 2\ha_\mu\hp_a \cr
& W_{ab}(\hp,\ha) =
\left[P^2 \delta_{ab} - m_1^2(\hp)\right] ({\hp_a\hp_b\over \hp^2})
+ \left[P^2 - m_2^2(\hp) - \xi\hp^2 \right]
(\delta_{ab} - {\hp_a\hp_b\over \hp^2}) \cr
& T_{ab}(\hp,\ha) =
\left[\gamma \cdot P - M_1(\hp) \right] \otimes I
- M_2(\hp)\otimes \sigma_3 - M_3(\hp)\otimes \sigma_1
+ \gamma \cdot A \otimes \sigma_2 \cr
& R(\hp) = \left[P^2 - \xi \hp^2\right]. \cr} \eqno\eq
$$
Here, $P_\mu$ is the momentum operator, and
$$
\eqalign{
& m_1^2(\hp) = m^2 + {\lambda\over 6} \hp^2 + {\tau\over 120} \hp^4 \cr
& m_2^2(\hp) = m^2 + {\lambda\over 2} \hp^2 + {\tau\over 24} \hp^4 \cr
& M_1(\hp) = M + g_1 \hp^2 \cr
& M_2(\hp) = g_2(\hp_1^2-\hp_2^2) \cr
& M_3(\hp) = 2g_2(\hp_1 \hp_2). \cr}\eqno\eq
$$

To one-loop order, the effective action is given by [\rJackiw]
$$
\Gamma[\hp,\ha]= {i\over 2} \tr\{\log W\}
+ {i\over 2} \tr\left\{\log(U-VW^{-1}V^{\dagger}) \right\}
+ {i\over 2} \tr\{\log T \} + i \tr\{\log R\}. \eqn\EfAc
$$
Note that the $1/2$ factor in the third term comes from the fact that
$\psi_a$ is a Majorana spinor.

On the other hand, there coluld exist in the effective action the following
parity-odd term [\rCHTh]
$$
\eqalign{
\Gamma^{odd}[\hp(x),\ha(x)] = \int d^3x\,\epsilon^{\mu\nu\rho}\biggl\{
& {\delta \kappa \over 2} \ha_\mu\partial_\nu \ha_\rho
+ C(\hp^2)\epsilon_{ab} \hp_a D_\mu\hp_b \partial_\nu \ha_\rho
+ \ldots \biggr\}. \cr}\eqn\WBCSI
$$
Let $\hp_a = \vp_a + \tvp_a(x).$  Expanding $\Gamma^{odd}[\hp(x),\ha(x)]$
around $\vp$ to linear order in $\tvp(x)$, we have
$$
\eqalign{
\Gamma^{odd}[\hp(x),\ha(x)] \approx \int d^3x\,\epsilon^{\mu\nu\rho}\biggl\{
& \bigl[{\delta \kappa \over 2} + \vp^2 C(\vp^2) \bigr]\ha_\mu\partial_\nu
\ha_\rho \cr
+ & {\partial \over \partial \vp^2}\bigl[\vp^2 C(\vp^2) \bigr]
\bigl[2(\vp \cdot \tvp) \ha_\mu\partial_\nu \ha_\rho \bigr] \cr
+ & C(\vp^2)
\bigl[\epsilon_{ab} \vp_a D_\mu \tvp_b \partial_\nu \ha_\rho \bigr]
\biggr\}. \cr}\eqn\WBCSII
$$
Therefore, to determine $\delta \kappa$, we must find out the coefficients
of the first two terms in the above equation.  Note that although the
coefficient of the third term is much easier to calculate, it is
unfortunately a total derivative term, and thus its coefficient can not be
uniquely determined.

Since we are interested in parity-odd part of the effective action, only the
second and the third terms in $\EfAc$ contribute.  Define
$$
X = U - V W^{-1} V^{\dagger}. \eqno\eq
$$
For the purpose of power counting, it is convenient to expand $X, U, V, W$
with respect to $\ha$
$$
\eqalign{X
& = X_0 + X_1 + X_2 + \ldots \cr
& = U_0 -(V_0+V_1)(W_0+W_1+W_2)^{-1}(V^{\dagger}_0+V^{\dagger}_1), \cr}
\eqno\eq
$$
where the subscripts denotes the powers of $\ha$.

Focusing on the relevant terms, we have
$$
\eqalign{
\Gamma^{odd}[\hp(x), \ha(x)] =
& {i\over2}\tr \biggl\{ [X^{-1}_0(\hp)]^{odd} X_2(\hp) \biggr\} \cr
- & {i\over2}\tr \biggl\{
[X^{-1}_0(\hp)]^{odd} X_1(\hp) [X^{-1}_0(\hp)]^{even} X_1(\hp) \biggr\} \cr
+ & {i\over4}\tr
\biggl\{T^{-1}_0(\hp) T_1(\hp) T^{-1}_0(\hp) T_1(\hp) \biggr\} + \ldots}
\eqn\WBCSIII
$$

To calculate the one-loop coefficient of the Chern-Simons terms,
we set $\hp_a = \vp_a$ in $\WBCSIII$.  Since $V_0 = 0$, we see $X_0 = U_0$
and the second term vanishes.  Consequently,
$$
\eqalign{
\Gamma^{odd}[\vp, \ha(x)] =
- & {i\over2}\tr \biggl\{
[U^{-1}_0(\vp)]^{odd} V_1(\vp) W^{-1}_0(\vp) V_1^{\dagger}(\vp) \biggr\} \cr
+ & {i\over4}\tr \biggl\{
T^{-1}_0(\vp) T_1(\vp) T^{-1}_0(\vp) T_1(\vp) \biggr\}, \cr}\eqn\CSCfI
$$
to one-loop order.  Here, in the Landau gauge
$$
\eqalign{
& [U_0^{-1}(\vp)]_{\mu\nu}
= {-e^2(p^2 - e^2 \vp^2) \left(\eta_{\mu\nu} - {p_\mu p_\nu \over p^2} \right)
-i \kappa e^4 \epsilon_{\mu\nu\rho}p^\rho \over
(p^2 - e^2\vp^2)^2 - \kappa^2 e^4 p^2}, \cr
& \hfill \cr
& [W_0^{-1}(\vp)]_{ab}
= {1\over p^2 - m_1^2(\vp)}\bigl( {\vp_a\vp_b \over \vp^2} \bigr)
+ {1\over p^2 - m_2^2(\vp)}
\bigl( \delta_{ab} - {\vp_a\vp_b \over \vp^2} \bigr), \cr
& \hfill \cr
& [T_0^{-1}(\vp)]
= { \bigl\{[\gamma\cdot p - M_1(\vp)]\otimes I + M_2(\vp)\otimes \sigma_3
+ M_3(\vp)\otimes \sigma_1 \bigr\} \over
[\gamma\cdot p - M_+(\vp)][\gamma\cdot p - M_-(\vp)]}, \cr}\eqn\Prop
$$
with
$$
M_\pm (\vp) = M + (g_1 \pm |g_2|)\vp^2.
$$

After some algebra, we have
$$
\eqalign{
\Gamma^{odd}[\vp, \ha(x)] =
- & {i\over2}\tr \biggl\{
\bigl[{-i \kappa e^4 \epsilon_{\mu\nu\rho} p^\rho \over
(p^2-e^2\vp^2)^2 - \kappa^2 e^4 p^2} \bigr]
\bigl[\ha^\mu\bigr] \bigl[{4\vp^2 \over p^2 - m_1^2(\vp) }\bigr]
\bigl[\ha^\mu\bigr] \biggr\} \cr
+ & {i\over2}\tr \biggl\{
\bigl[ {1 \over \gamma\cdot p - M_+(\vp)} \bigr] [\gamma\cdot \ha]
\bigl[ {1 \over \gamma\cdot p - M_-(\vp)} \bigr] [\gamma\cdot \ha] \biggr\}.
\cr}\eqn\CSCfII
$$
Employing the technique of derivative expansion and Wick rotation [\rDeri],
one can see the bosonic part contributes
$$
{4\over 3} \int {d^3p\over(2\pi)^3}
{\kappa e^4 \vp^2 p^2\over\left[ (p^2+e^2\vp^2)^2 + \kappa^2 e^4 p^2\right]
\left[p^2+m_1^2(\vp)\right]^2}\eqn\CSCfB
$$
to the coefficient of the Chern-Simons term.  In the pure Chern-Simons limit
\hfill\break
$e \to \infty,$ it gives
$$
{m(2 |m| + |m_1|) \over 6\pi (|m| + |m_1|)^2}, \eqn\RB
$$
where $m = \vp^2/ \kappa$.  When $m_1 = m$, the above result reduces
to $m/(8\pi |m|)$.

Similarly, we see the fermionic part contributes
$$
-{2\over 3} \int {d^3p\over(2\pi)^3}
{\left[M_+(\vp) + M_-(\vp)\right]\left[p^2 + M_+(\vp) M_-(\vp)\right] \over
\left[(p^2+M_+^2(\vp) \right]^2 \left[p^2+M_-^2(\vp)\right]^2}.\eqn\CSCfF
$$
After integration, it leads to the following expression
$$
{-(M_+ + 2M_-)|M_+| -(2M_+ + M_-)|M_-| \over 12\pi (|M_+| + |M_-|)^2}.
\eqn\RF
$$
Note that the results in Eqs. $\RB$ and $\RF$ are identical to those obtained
in [\rCscf].  For the special case $M_+ = M_-$,  we have a Dirac fermion with
mass $M_+$ and the above result simplifies to $-M_+/(8\pi |M_+|)$.  In
contrast when $M_+ = -M_-$, we have two Majprana fermions with opposite spin
and their contributions exactly cancel out.

{}From $\WBCSII$, it is easy to see
$$
{\delta \kappa \over 2} + \vp^2 C(\vp^2) =
{4 \over 3}I_B(\vp^2) - {2\over 3} I_F(\vp^2), \eqn\CSCfIII
$$
where $I_B(\vp^2)$ and $I_F(\vp^2)$ are the integrations in Eq.$\CSCfB$ and
$\CSCfF$.

To further determine $C(\vp^2)$, we must evaluate $\WBCSIII$ to linear order
in $\tvp$.  It is easy to see that the second term still has no contribution
and
$$
\eqalign{
\Gamma^{odd}[\vp + \tvp(x), \ha(x)] =
& {i\over2}\tr \biggl\{ \bigl[
U^{-1}_0(\vp) \tvp_a {\partial U_0(\vp)\over \partial \vp_a} U^{-1}_0(\vp)
\bigr]^{odd} V_1(\vp) W^{-1}_0(\vp) V_1^{\dagger}(\vp) \biggr\} \cr
+ & {i\over2}\tr \biggl\{ \bigl[U^{-1}_0(\vp) \bigr]^{odd} V_1(\vp) \bigl[
W^{-1}_0(\vp) \tvp_a {\partial W_0(\vp)\over \partial \vp_a} W^{-1}_0(\vp)
\bigr] V_1^{\dagger}(\vp) \biggr\} \cr
- & {i\over2}\tr \biggl\{ T^{-1}_0(\vp) T_1(\vp) \bigl[
T^{-1}_0(\vp) \tvp_a {\partial T_0(\vp)\over \partial \vp_a} T^{-1}_0(\vp)
\bigr] T_1(\vp) \biggr\} + \ldots \cr}\eqn\CSCfIV
$$

After tedious but straightfoward calculation, we obtain the coefficient for
the second term in $\WBCSII$:
$$
{\partial \over \partial \vp^2}
\biggl\{{4 \over 3}I_B(\vp^2) - {2 \over 3} I_F(\vp^2) \biggr\}. \eqno\eq
$$
Imposing the boundary condition that
$\bigl[ \vp^2 C(\vp^2) \bigr] \big|_{\vp =0} = 0,$ we have
$$
\vp^2 C(\vp^2) = {4 \over 3}I_B(\vp^2) - {2 \over 3} I_F(\vp^2)
+ {M \over 8\pi |M|}. \eqno\eq
$$
{}From Eq.$\CSCfIII$, it is easy to see that
$$
\delta \kappa = - {M \over 4\pi |M|}. \eqno\eq
$$
In other words, the one-loop corrections to $\kappa$ in the symmetric and
asymmetric phases are the same, if we subtract out the contrbution from the
would be Chern-Simons term.

Let us now apply the above results to the $N=2$ and $N=3$ self-dual
Chern-Simons Higgs systems [\rCSII, \rCSIII].  For the $N=3$ case, there
are two Dirac fermions $(\psi, \chi)$ with mass $m/2$ and $-m/2$ in the
symmetric phase.  In the Higgs phase, the gauge boson and the Higgs field
have the same mass $m$.  Moreover, the mass of $\psi$ flips sign while
$\chi$ splits into two Majorana fermions with mass of opposite sign.
Hence, we have
$$
\eqalign{
& \delta \kappa \big |_{N=3} =  0 \cr
& \biggl[\vp^2 C(\vp^2) \biggr]_{N=3}
=  {1\over 2\pi} {\kappa \over |\kappa|}. \cr}\eqno\eq
$$
The $N=2$ result can be obtained by discarding the contribution from $\chi$
and
$$
\eqalign{
& \delta \kappa \big |_{N=2} =  {-1\over 2\pi} {\kappa \over |\kappa|} \cr
& \biggl[\vp^2 C(\vp^2) \biggr]_{N=2}
=  {3\over 4\pi} {\kappa \over |\kappa|}. \cr}\eqno\eq
$$
In view of $\WBCSI$, we see the quantization of of the corrections to the
Chern-Simons coefficient in these systems is a reflection of the quantization
of the anomolous magnetic moment of the charged scalars.

In this letter, we have extended the result in [\rCHTh] and shown that the
one-loop correction to the Chern-Simons
coefficient in the Higgs phase is identical to that in the symmetric phase and
therefore originates only from the fermionic part, if we properly remove the
contribution from the would be Chern-Simons term.  An interesting question is
whether the Coleman Hill theorem restated in terms of the effective action
holds to all loops.  As for the nonabelian case, our naive expectation is that
after we subtract out the contribution from the would be Chern-Simons term,
the correction to the Chern-Simons coefficient obtained from evaluating the
vacuum polarizations of the broken and unbroken gauge bosons are identical
and quantized [\rDunne, \rKhare], in particular the $SU(2)$ case even though
the gauge symmetry is completely broken there.  However, more detailed
analysis shows that this is not the case if the the remaining symmetry group
is not simple [\rKhare].  Finally, since the would be Chern-Simons term is
invariant even under the large gauge transformation, these theoriers are
quantum-mechanically consistent.

\vskip 0.2in

\centerline{\bf Acknowledgement }

I would like to thank K. Lee for helpful comments and criticisms, as well as
M. Paranjape for a critical reading of the manuscript.  This work is supported
National Science Council, Taiwan (Grant No. NSC84-2112-M-001-022).

\vfill \endpage

\refout

\end